# Reduction in turbulence-induced non-linear dynamic vibration using tuned liquid damper (TLD)

**Ananya Majumdar[1], Biplab Ranjan Adhikary[1] and Partha Bhattacharya[1]**

[1]Department of Civil Engineering, Jadavpur University, Kolkata-700032, India

**ABSTRACT**

In the present research work, an attempt is made to develop a coupled non-linear turbulence-structure-damper model in a finite volume-finite difference (FV-FD) framework. Tuned liquid damper (TLD) is used as the additional damping system along with inherent structural damping. Real-time simulation of flow-excited bridge box girder or chimney section and the vibration reduction using TLD can be performed using the developed model. The turbulent flow field around a structure is modeled using an OpenFOAM transient PISO solver, and the time-varying drag force is calculated. This force perturbs the structure, causing the sloshing phenomena of the attached TLD, modeled using shallow depth approximation, damping the flow-induced vibration of the structure. The structural motion with and without the attached TLD is modeled involving the FD-based Newmark-Beta method using in-house MATLAB codes. The TLD is tuned with the vortex-shedding frequency of the low-Reynolds number flows, and it is found to be reducing the structural excitation significantly. On the other hand, the high-Reynolds number turbulent flow exhibits a broadband excitation, for which by tuning the TLD with few frequencies obtained through investigations, a good reduction in vibration is observed.

## 1. INTRODUCTION

Turbulence proves its' prominent presence in our surroundings, which can adversely affect the important structures like bridges, high-rise buildings or chimneys. The excessive flow-induced vibration even leads to collapse. One of such an iconic example is Tacoma bridge collapse and a very recent, the under-construction bridge collapse in Sultangunj, India on 29th April 2022.

In order to reduce the turbulence-excited structural motion it is important to properly model the flow field surrounding the structure. Next, the structural motion is to be estimated with and without a TLD attached to it.

To obtain the desired reduction in the excitation level of the structure, several active and passive damping techniques are used by researchers and engineers. One of the widely used passive damping instruments is TLD, which is essentially tuned to the first fundamental frequency [1,2] or first two fundamental frequencies [3] of the vibrating structure for the best damping experience. In the case of harmonic excitation and non-deterministic forcing with a prominent frequency value, the TLD is tuned to the excitation frequency [4]. In case of turbulent flow, the behavior of TLD and possible tuning frequency is not studied yet.

## 2. LITERATURE REVIEW AND OBJECTIVE

Many researchers have investigated the flow past bluff body for different Reynolds numbers (Re) using numerical [5] or experimental techniques [6-8]. The Strouhal number (St) and drag coefficient ($C_D$) are evaluated and reported. Structural excitation due to any random forcing is estimated using the FD-based non-linear Newmark-Beta method in time domain [4]. The sloshing behavior of TLD and resulting structural response reduction is estimated numerically and/or experimentally [1-4] using harmonic or random earthquake ground acceleration.

However, the coupled non-linear model to capture the turbulence-induced structural motion and its reduction using TLD is non-existent.

Therefore, in the present research work, an attempt is made to develop a numerical FV-FD-based non-linear model that will estimate the turbulence forcing, and simultaneously at each time step, it will capture the sloshing-induced base shear produced by the TLD liquid at the TLD-structure junction, eventually producing additional damping to the SDOF system.

## 3. METHODOLOGY

The entire work is subdivided into *two* subsections, each consisting of modelling a part of the full solution technique.
   a) Modelling the turbulent flow-field past a rigid obstacle and estimation of the drag force applied to it.
   b) Using the time-varying turbulent forcing to estimate the structural response at each time step with and without attached TLD.

### 3.1 Turbulence past a rigid obstacle

Unsteady or transient simulation for isotropic turbulence is performed using Large Eddy Simulation (LES), which essentially works based on filter operation. Eddies larger than a certain length scale, typically in the order of the grid size, are fully resolved. The smaller eddies are modelled using a sub-grid scale model. As the turbulence is considered to be isotropic, only the size of these smaller eddies becomes important, not the shape. Once the mean velocity field is computed by the RANS model, the fluctuating velocity component ($u'$) can conceptually be estimated by subtracting time-averaged mean velocity $\bar{U}$, from instantaneous velocity, $U$. This fluctuating velocity is then used to calculate the turbulent kinetic energy (k) per unit mass, as $k = \frac{1}{2} u' u'$. If all the fluctuating components in three directions are considered, the Reynolds stress tensor can be estimated (per unit density) in symmetric matrix form, and in 3D and 2D domains the resolved kinetic energy becomes

$$k_{res} = \frac{1}{2}[\overline{(u')^2} + \overline{(v')^2} + \overline{(w')^2}] \quad (1)$$

$$k_{res} = \frac{1}{2}[\overline{(u')^2} + \overline{(v')^2}] \quad (2)$$

The amount of the remaining kinetic energy is termed sub-grid scale kinetic energy, $k_{sgs}$, and calculated by a sub-grid eddy viscosity model. In the present study, the Smagorinsky-Lilly model is used, where an additional stress term ($\tau_{sgs}$) is applied to break down the eddies larger than the mesh size because molecular viscosity is not sufficiently strong to do so. This stress term can be derived by applying filtering operation on compressible Navier-Stokes equations as follows:

$$\frac{\partial \rho}{\partial t} + \frac{\partial (\rho U_j)}{\partial x_j} = 0 \quad (3)$$

$$\frac{\partial (\rho U_i)}{\partial t} + \frac{\partial (\rho U_i U_j)}{\partial x_j} = -\frac{\partial P}{\partial x_i} + \frac{\partial}{\partial x_j}(\tau_{ij} + \tau_{sgs}) \quad (4)$$

This sub-grid stress $\tau_{sgs}$ is modelled using Eq. (5)

$$\tau_{sgs} = 2\rho v_{sgs} S_{ij}^* - \frac{2}{3}\rho k_{sgs}\delta_{ij} \quad (5)$$

$$S_{ij}^* = \frac{1}{2}\left(\frac{\partial U_i}{\partial x_j} + \frac{\partial U_j}{\partial x_i} - \frac{1}{3}\frac{\partial U_k}{\partial x_k}\partial_{ij}\right) \quad (6)$$

Assuming the profile to be linear within the viscous sublayer, and obeying the 1/7$^{th}$ power law in the outside region, the wall function is formulated as described by Germano *et al.* (1991).

$$U^+(y^+) = \begin{cases} y^+ & if\ y^+ < 11.8 \\ 8.3(y^+)^{1/7} & if\ y^+ > 11.8 \end{cases} \quad (7)$$

where non-dimensionalized velocity ($U^+$) and wall distance ($y^+$) are given as

$$U^+ = \frac{U}{U_\tau}, y^+ = \frac{U_\tau y}{\nu} \quad (8)$$

Lilly (1966) proposed a value for the Smagorinsky constant $C_s$ as 0.173, considering turbulence to be homogeneous and isotropic, which is true for a shear-free turbulent event far from any wall. However, modern CFD codes typically use different values of $C_s$ for near-wall turbulent events. In the present wall-function approach (SLWF) Smagorinsky-Lilly constant, $C_s$ is considered as 0.1 to determine the sub-grid scale kinematic eddy viscosity.

In the Pressure-Implicit with Splitting of Operators (PISO) algorithm, steady-state flow problems can be solved using the LES model with a pressure corrector.

The coefficient of drag is given by

$$C_D = \frac{F_D}{\frac{1}{2}\rho U^2 A} \quad (9)$$

$F_D$: Drag Force, A: Projected area

### 3.2 Sloshing of tuned liquid damper (TLD)

At any point wave height is h. Assuming shallow wave theory to be valid and no point of time wave is reaching the tank top, partial differential equations for the sloshing motion of liquid can be written as [4],

$$\frac{\partial h}{\partial t} + h\frac{\partial v}{\partial x} + v\frac{\partial h}{\partial x} = 0 - \left(\frac{1}{2\beta} - 1\right)\ddot{u}_n \quad (10)$$

$$\frac{\partial v}{\partial t} + g\frac{\partial h}{\partial x} + v\frac{\partial v}{\partial x} - g(\theta_{TLD} - S) + \frac{\partial^2 v}{\partial t^2} = 0 \quad (11)$$

Where, u is the displacement of the structure at any point of time, thus, displacement of the TLD liquid surface, as it is attached to the structure. $\theta_{TLD}$ is the rotational displacement, h is the wave height at location x and time t, v is the particle velocity at location x and time t. The boundary conditions are given by:

$$v(0,t) = v(L,t) = 0 \quad (12)$$

Initial conditions: At the starting of time (t = 0) steady state condition is considered,

$$h(x,0) = h_0;\ v(x.0) = 0 \quad (13)$$

Non-dimensional slope of the energy gradient line is written as,

$$S = \frac{\tau_w}{\rho g h} \quad (14)$$

The wall shear stress at the base of the tank, $\tau_w$ is given by,

$$\tau_w = \frac{\mu_f v_{max}}{h}, for\ z \leq 0.7 \quad (15)$$

$$\tau_w = \sqrt{\rho \mu_f \omega}, for\ z > 0.7 \quad (16)$$

$\omega$ is the forcing frequency, $z = \sqrt{\frac{\omega g}{2\mu_f}}$; $\mu_f$ is the absolute viscosity of fluid, here water. Sloshing force (F) acting on the walls of the water tank is given by:

$$F = 0.5\rho g B(h_R^2 - h_L^2) + \int_0^L \rho g B h S dx \quad (17)$$

$h_R, h_L$: Wave height w.r.t bottom of the tank at the right and left wall of the tank. $\rho$: Density of fluid, here water.
The sloshing frequency of the tank liquid is given as,

$$f_w = \frac{1}{2\pi}\sqrt{\frac{\pi g \tanh(\frac{\pi h}{L})}{L}} \quad (18)$$

$f_w$ = Fundamental natural frequency of liquid sloshing
L = Tank length

### 3.3 Solving non-linear partial differential equations

The non-linear partial differential equations used to describe tuned liquid damper system discussed in the previous section are solved in iterative finite difference technique. Any general function f can be written as,

$$f = \alpha f_i + (1-\alpha)\frac{f_{i+1} + f_{i-1}}{2} \quad (19)$$

The function can be velocity (v), wave height (h), or slope of the energy gradient line (S). In this numerical approach space derivatives are estimated as,

$$\frac{\partial f}{\partial x} = \frac{f_{i+1} - f_{i-1}}{2\Delta x} \quad (20)$$

for a particular time step. $\Delta x$ is the element dimension. 'i' is the node number. Time derivatives are estimated as,

$$\frac{\partial f}{\partial t} = \frac{f_i^{k+1} - \left[\alpha f_i + (1-\alpha)\frac{f_{i+1}+f_{i-1}}{2}\right]}{\Delta t} \quad (21)$$

$\Delta t$ is the time step size. 'k' is the time step number. $\alpha = 0.98$

### 3.4 Newmark-β Method

In this method response of a damped or undamped structural system can be evaluated by involving numerical discretization of space and time. The second-order differential equation for a general excited structural system can be written as,

$$M\ddot{u} + C\dot{u} + Ku = F_{turbulence} \quad (22)$$

Using the extended mean value theorem, the first time-derivative of the motion is expressed as

$$\dot{u}_{n+1} = \dot{u}_n + \Delta t \ddot{u}_\alpha \quad (23)$$

Where, $\ddot{U}_\alpha = (1-\alpha)\ddot{U}_n + \alpha \ddot{U}_{n+1}, 0 \leq \alpha \leq 1$ (24)

This transforms the Eq. 23 into

$$\dot{u}_{n+1} = \dot{u}_n + (1-\alpha)\Delta t \ddot{u}_n + \alpha \Delta t \ddot{u}_{n+1} \quad (25)$$

As to incorporate the temporal change of the second time derivative of the motion, in the estimation of the displacement, the following formulations are suggested,

$$u_{n+1} = u_n + \Delta t \dot{u}_n + \frac{1}{2}\Delta t^2 \ddot{u}_\beta \quad (26)$$

Where, $\ddot{u}_\beta = (1-2\beta)u_n + 2\beta \ddot{u}_{n+1}, 0 \leq 2\beta \leq 1$ (27)

Therefore, Eq. 26 becomes

$$u_{n+1} = u_n + \Delta t \dot{u}_n + \frac{1}{2}(1-2\beta)\Delta t^2 \ddot{u}_n + \beta \Delta t^2 \ddot{u}_{n+1} \quad (28)$$

The finite difference formulas for the Newmark Beta scheme are

$$\ddot{u}_{n+1} = \frac{1}{\beta \Delta t^2}(u_{n+1} - u_n) - \frac{1}{\beta \Delta t}\dot{u}_n - \left(\frac{1}{2\beta} - 1\right)\ddot{u}_n \quad (29)$$

$$\dot{u}_{n+1} = \frac{1}{\beta \Delta t}(u_{n+1} - u_n) - \left(\frac{\alpha}{\beta} - 1\right)\dot{u}_n - \Delta t\left(\frac{\alpha}{2\beta} - 1\right)\ddot{u}_n \quad (30)$$

## 4. RESULTS AND DISCUSSION

### 4.1 Validation 1: Suitability of SLWF model

A flow domain of 21D height is considered for the problem, where, D is the diameter of the cylinder. Upstream and downstream boundary distances of the domain are kept at 8.5D and 20.5D from the center of the cylinder, respectively. D is taken as 19mm or 0.019m. Inlet flow velocity is kept uniform with a magnitude of 69.2 m/s, which leads to a flow Reynolds number of 90,000. An attempt has been made to numerically replicate the experiment performed by Revell et al. [12]. A schematic of the problem is presented in Fig. 1.

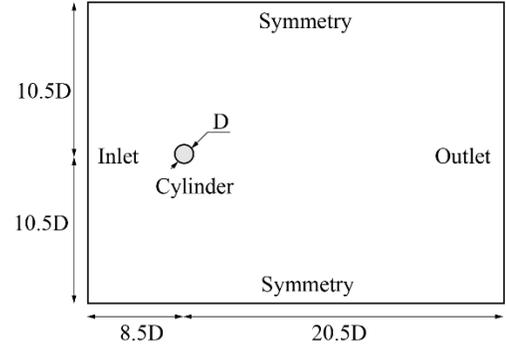

**Fig 1. Schematic of flow past bluff body**

At first, transient simulation (LES) is performed to extract globally averaged flow parameters, like wall shear stress, $\tau_w$ coefficient of skin friction $C_f$, etc. at different points on the cylinder wall, using both OpenFOAM (v-2012) and ANSYS-Fluent (V14.5). The mesh refinement is optimized as suggested by Wang et al. [10]. A total of 94,651 quadrilateral cells are used. An unstructured grid with necessary refinement close to the cylinder wall is used in the form of 360 and 80 mesh nodes in the circumferential and radial directions, respectively. Incompressible Navier-Stokes equations are solved at each grid in a PISO solver. To reduce the computation cost, the Non-Iterative Time-Advancement scheme is chosen along with the Fractional Step method (FSM). Spatial discretization is performed using the Bounded Central Difference scheme. Unsteady pressure is interpolated over the entire domain using PRESTRO algorithm. To keep Courant-Friedrichs-Lewy (CFL) number below 1.5, the time step size is chosen to be $1 \times 10^{-6}$ s. Two different sub-grid eddy viscosity models are used for computation, a) the Smagorinsky-Lilly model with Wall Function (SLWF) and b) the Dynamic Smagorinsky-Lilly model (DSL). The comparison results obtained from ANSYS-Fluent is presented in Fig 2 and Fig 3 in order to select the suitable model between SLWF and DSL.

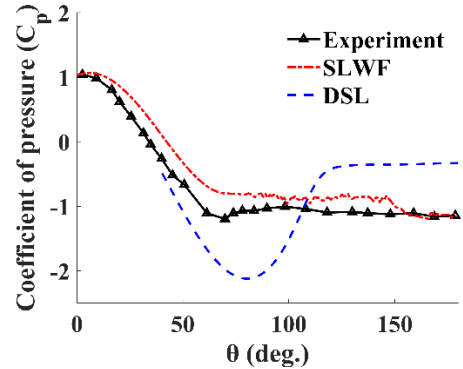

**Fig 2. Coefficient of pressure at different angle ($\theta$) of the points over the body, measured from the trailing edge. (Experimental results are from Revell et al. [12])**

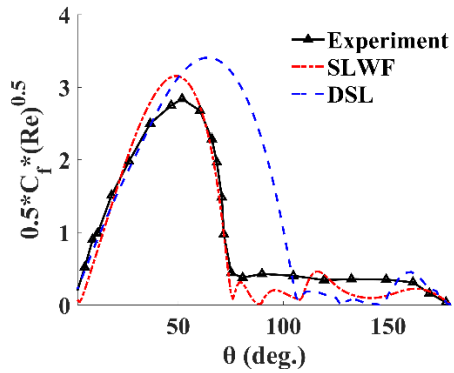

**Fig 3.** Scaled skin friction coefficient at different angle (θ) of the points over the body, measured from the trailing edge. (Experimental results from Revell et al. [12])

Between these two models, SLWF tends to behave better for flow past bluff body, that is why, the next problem case of flow past square cylinder for high Re is evaluated with this method.

**4.2 Validation 2: Flow past square cylinder**

A square cylinder of dimension D, is placed in a fluid domain, with a free stream velocity of $U_0$ at zero angle of attack. An inlet turbulence of intensity 0% is set at first, which is the laminar approaching flow case [5]. From this case, the mean and root mean square values of coefficient of drag ($C_D$), have been evaluated and compared with the experimental results [6-8], and presented in Table 1. The schematic of the general problem is shown in Fig. 4, and meshing in Fig. 5. For the present validation purpose, TLD is not considered.

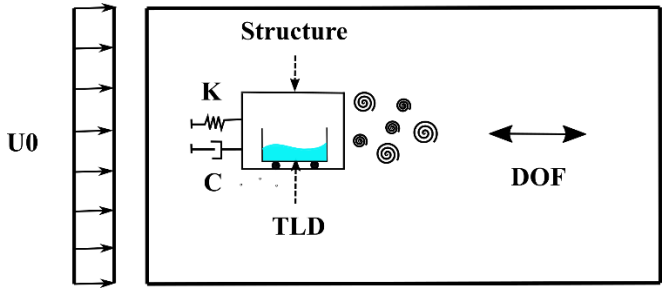

**Fig 4.** Schematic of the turbulence-structure-TLD system

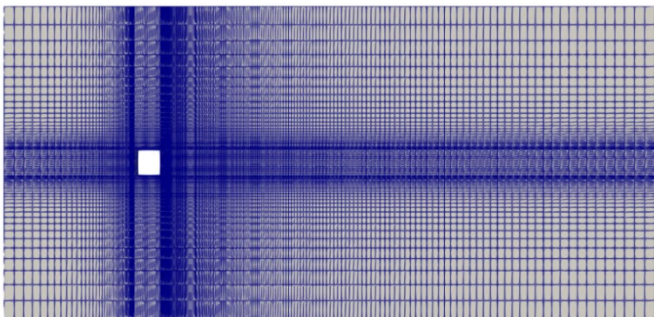

**Fig 5.** Meshing in OpenFOAM

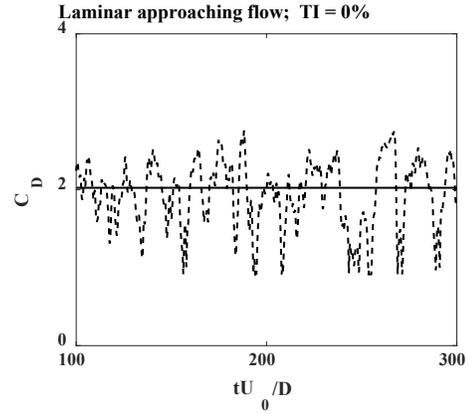

**Fig 6.** Coefficient of drag over time

**Table 1:** Validation for flow past square cylinder

|  | Re | St | $\overline{C_D}$ | $\acute{C}_D$ |
|---|---|---|---|---|
| Experimental (Lee 1975) [6] | $1.76 \times 10^5$ | 0.122 | 2.07 |  |
| Experimental (Pocha 1971) [7] | $9.1 \times 10^4$ | 0.12 | 2.06 | 0.19 |
| Experimental (Noda and Nakayama 2003) [8] | $6.89 \times 10^4$ | 0.131 | 2.16 | 0.207 |
| Numerical (Li et. al 2018) [5] | $10^5$ | 0.129 | 2.085 | 0.218 |
| Present study | $10^5$ | 0.101 | 2.025 | 0.206 |

As seen in Fig. 6 the turbulent flow is simulated for an initial smaller time duration, which possibly resulted in a lower $C_D$ value. A better prediction is expected for a longer simulation.

**4.3 Validation 3: TLD-structure interaction**

In order to validate the developed TLD-structure interaction model, the shaking table experiment performed by Sun et al. [2] is considered. TLD tank is 59cm long, and 33.5cm wide. The water depth is 3cm. The frequency ratio is the ratio between the natural frequency of the structure to that of the TLD liquid (here water). The structural displacement with and without attached TLD for different frequency ratios is presented in Fig. 7. The frequency ratio vs maximum and minimum wave height is presented in Fig. 8.

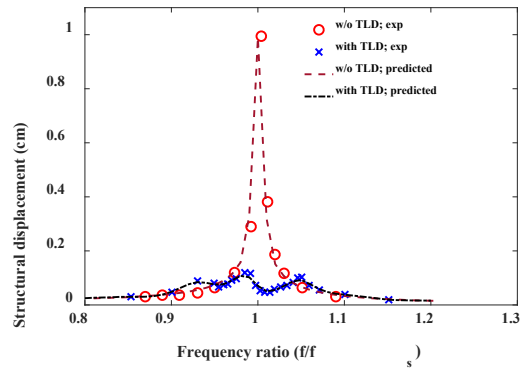

**Fig 7.** Structural displacement with and without TLD [2]

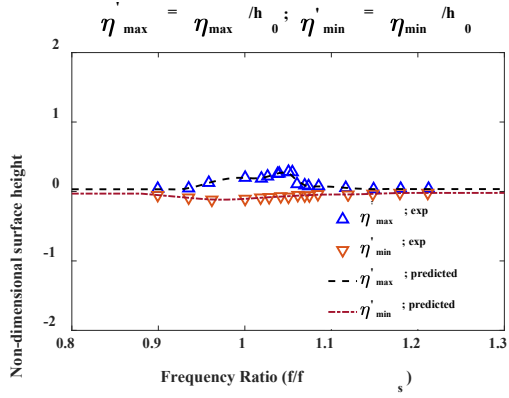

Fig 8. Maximum and minimum surface height [2]

Once the turbulence modelling around a bluff body and the TLD-structure interaction modelling are validated, next two case studies are performed for laminar (Re = 250) and turbulent (Re = $10^5$) flows. In both the cases time-varying drag force is used as the forcing to the next stage of TLD-structure interaction model. The general problem schematic is presented earlier in Fig. 1. The obstacle is considered as the square cylinder used in the validation 2 problem (section 4.2). The results from the two problems are presented and discussed in sections 4.4 and 4.5.

### 4.4 Turbulence-structure-TLD interaction; Re = 250

In this segment, a laminar flow (Re = 250) is considered. The computed drag coefficient and Strouhal numbers are presented in Table 2. This drag coefficient is subsequently used to estimate the force and the TLD parameters are so chosen that it is tuned to the corresponding natural frequency of 0.538Hz (for St = 0.138). The predicted damped and undamped displacement (non-dimensionalized by the cylinder dimension, D) is presented in Fig. 8. The inherent structural damping is taken as 0.32% [2]. The spring stiffness is calculated accordingly.

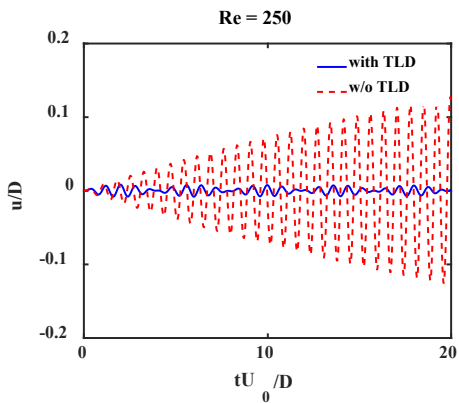

Fig 9. Structural response; natural frequency of 0.538Hz;

Table 2: Drag coefficient for Re = 250 flow case

| Numerical (Franke 1990) [13] | Re | St | $C_D$ |
|---|---|---|---|
|  | 250 | 0.141 | 1.67 |
| Present study | 250 | 0.138 | 1.78 |

### 4.5 Turbulence-structure-TLD interaction; Re = $10^5$

In this segment, a turbulent flow (Re = $10^5$) is considered. The computed drag coefficient is already presented in Table 1. This drag coefficient is subsequently used to estimate the force.

As the turbulent flow produces a broadband spectrum it is very difficult to find any particular forcing frequency, and eventually tuning the TLD also becomes challenging. In order to estimate the TLD performance in the reduction of the structural response, an extensive study is carried out in two stages.

a) TLD properties are varied to obtain different natural frequencies, replicating a broadband spectrum (using Eq. 18)
b) For each frequency the TLD-structure interaction is simulated and the performance is measured in terms of the ratio of the maximum displacement with and without TLD.

The result of the performance study is presented in Fig. 10.

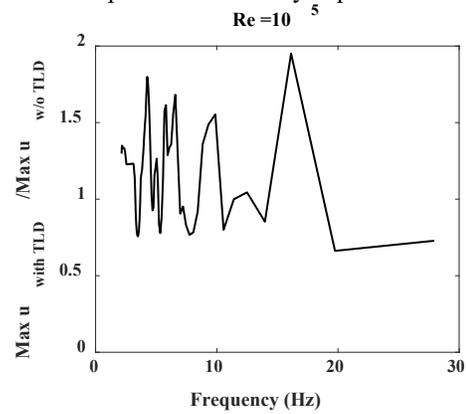

Fig 10. Performance of TLD in a broadband spectrum

From this study, few frequencies are identified where the TLD can reduce the structural response. In the other frequencies, the effect of attached TLD is found to be adverse though. Next, the structural response with and without TLD at a few identified frequencies are presented in Fig. 11 and Fig. 12.
The predicted damped and undamped displacements are non-dimensionalized by the cylinder dimension, D.

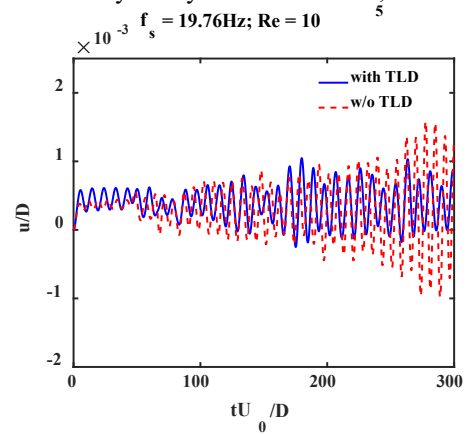

Fig 11. TLD performance at 19.76Hz frequency

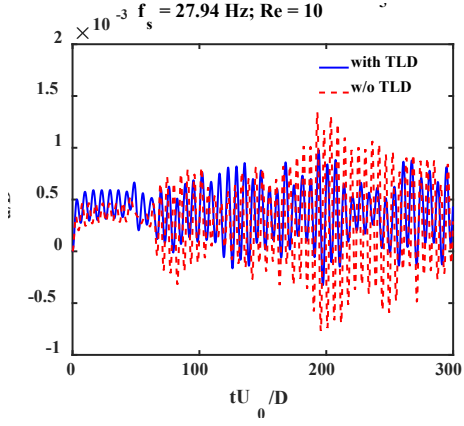

**Fig 12. TLD performance at 27.94Hz frequency**

## 5. CONCLUSIONS

In the present research work, a turbulence-structure TLD model is proposed, developed in FD framework. Open-source flow solver (OpenFOAM) and in-house MATLAB based TLD- structure interaction model is used. The significant conclusions are:

a) In case of laminar flow induced structural vibration, the TLD works perfectly when it is tuned with the vortex-shedding frequency.
b) In case of turbulence, the broadband force spectrum restricts to select any particular tuning frequency. However, in the present work a detailed study is carried out to identify few possible tuning frequencies that eventually is found to be reducing the structural response for the particular present case.
c) There is a requirement of further extensive studies with different turbulent conditions involving two-way FSI model to reach at any practical conclusive decision.

## NOMENCLATURE

| | | |
|---|---|---|
| $U$ | Turbulent velocity | [m/s] |
| $u$ | Structural displacement | [m] |
| $\dot{u}$ | Structural velocity | [m/s] |
| $\ddot{u}$ | Structural acceleration | [m/s$^2$] |
| $v$ | TLD liquid velocity | [m/s] |
| $u'$ | fluctuating velocity component | [m/s] |
| $\overline{U}$ | time-averaged mean velocity | [m/s] |
| $k$ | turbulent kinetic energy per unit mass | [J/kg] |
| $k_{sgs}$ | sub-grid scale kinetic energy | [J] |
| $\tau_{sgs}$ | sub-grid stress | [N/m$^2$] |
| $C_D$ | Drag coefficient | -- |
| $\overline{C_D}$ | Mean Drag Coefficient | -- |
| $\acute{C}_D$ | Root mean square drag coefficient | -- |
| $F_D$ | Drag Force | [N] |
| $A$ | Projected area | [m$^2$] |
| $h$ | Wave height at a particular location x | [m] |
| $t$ | Time | [s] |
| $\mu_f$ | Absolute viscosity of water | [Pa-s] |
| $\rho_f$ | Density of water | [kg/m$^3$] |
| $L$ | Tank length | [m] |
| $\rho$ | Density of air | [kg/m$^3$] |
| $F$ | Sloshing force acting on the walls | [N] |